# Enhanced light emission from gap plasmons in nano-strip MIM tunnel junctions

**Saurabh Kishen, Jinal Tapar and Emani Naresh Kumar**

[1] Department of Electrical Engineering, Indian Institute of Technology, Hyderabad, 502285, India.

E-mail: nke@iith.ac.in

**Abstract**

Electrical excitation of light using inelastic electron tunneling is a promising approach for the realization of ultra-compact on-chip optical sources with high modulation bandwidth. However, the practical implementation of these nanoscale light sources presents a challenge due to the low electron-to-photon transduction efficiencies. Here, we investigate designs for the enhancement of light generation and out-coupling in a periodic Ag-SiO$_2$-Ag tunnel junction due to inelastic electron tunneling. The structure presents a unique advantage of a simple fabrication procedure as compared to the other reported structures. We achieve a resonant enhancement in the local density of optical states up to three orders of magnitude over vacuum for the periodic metal-insulator-metal tunnel junction. By efficiently coupling the gap plasmon mode and the lattice resonance, an enhanced radiative efficiency of ~0.53 was observed, 30% higher as compared to the uncoupled structure.

Keywords: gap plasmon, inelastic electron tunneling, lattice resonance

## 1. Introduction

Following its initial demonstration in 1976 [1], the study of electrical excitation of surface plasmon (SP) through the process of inelastic electron tunneling (IET) has expanded rapidly in recent years [2-11]. IET based devices have a unique potential to create ultra-compact photon and SP sources, and can be readily integrated with existing CMOS technologies for practical on-chip integration. The metal-insulator-metal (MIM) tunnel junction is a particularly promising candidate for demonstrating IET based nanoscale optical devices due to its ability to confine optical mode into deep sub-wavelength volumes, while the metal cladding can serve as an electrical contact. The emission through IET is a two-stage process: electrons tunneling inelastically through a MIM tunnel junction lose a part of their energy in the barrier to the gap plasmon modes of energy $\hbar\omega$, provided the applied bias $|eV_b| \geq \hbar\omega$; the gap plasmon mode then decays to emit either photons [5-8] or propagating SPs [9-11].

Though the IET based devices are known for ultra-small device footprint and ultra-large modulation bandwidth, they exhibit low external quantum efficiencies (EQE) in the range of ~2% for both electron-to-photon and electron-to-surface plasmon conversion [12]. The low efficiencies for light generation can be attributed primarily to two factors: (a) the elastic tunneling rate is substantially higher than the inelastic tunneling rate leading to lower electron-to-gap plasmon coupling efficiency and, (b) the low radiative out-coupling of the excited gap plasmons to free-space photons. The former can be enhanced by designing a structure with high local density of optical states (LDOS) for gap plasmons within the barrier [13], and the latter can be overcome by using a plasmonic antenna with high radiative efficiencies [14, 15]. Here, we theoretically and numerically investigate the resonant enhancement in total LDOS and the radiative efficiency of gap plasmons in a 2D periodic Ag-SiO$_2$-Ag structure with a barrier thickness of 3 nm. The advantage of the structure stems from the ease of fabrication as compared to other reported structures [16]. The top-down nanofabrication approach can be employed in conjunction with focused-ion beam milling for realizing the patterned MIM tunnel junction. The use Ag-SiO$_2$ ensures that the probability of the tunneling electrons coupling to the gap plasmon mode is significantly higher than the quenching rate at the wavelength of interest [17]. In this paper, we report the impact of structural parameters on the radiative efficiency of the MIM tunnel junctions. For a particular periodicity, the highest radiative efficiency at the desired wavelength is achieved by tuning the width and height of the top electrode. By varying the periodicity, the out-coupling efficiency of the gap plasmon is further improved by ~30%, resulting from the strong coupling between the gap mode and the lattice resonance.

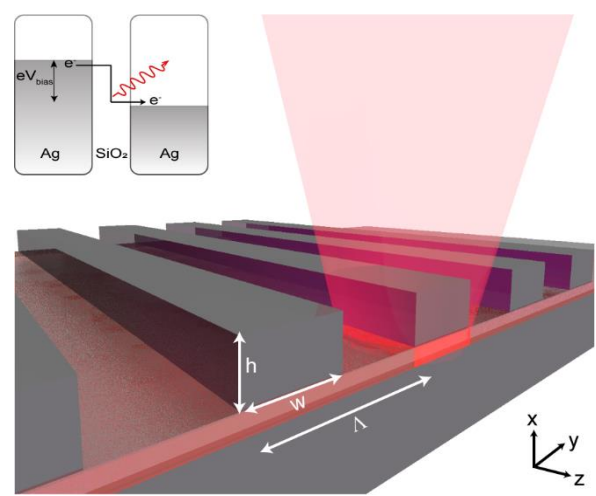

Figure 1. Simulated device consists of a periodic silver nano-strip antenna ($\Lambda$ = 400 nm) of $w$ = 35 nm and $h$ = 80 nm on a silver film of 200 nm thickness with a 3 nm thick SiO$_2$ sandwiched in between. The inset depicts the generation of photons through inelastic electron tunneling when $V_{bias}$ is applied across the tunnel junction.

## 2. Modelling and Simulation

The simulated structure depicted in Fig. 1 consists of a continuous SiO$_2$ film of thickness 3 nm sandwiched between a continuous silver bottom film of thickness 200 nm and periodic silver nano-strip of width $w$ and height $h$. The nano-strip acts as the top electrode, and the silver film as the bottom electrode for the tunnel junction. The structure is periodic with a periodicity ($\Lambda$) of 400 nm in the $y$ direction, and semi-infinite in the $z$ direction. On the application of bias voltage, photons are emitted from the junction as a result of quantum-mechanical tunneling (see inset of Fig. 1). The efficiency of IET as a source of gap plasmon can be described using Fermi's golden rule, and it depends on the applied bias and the total LDOS. The total light emission then depends on the efficiency with which the antenna radiates out the gap plasmon mode, i.e. the radiative efficiency of the antenna [5]. The spectrum of emitted radiation is modelled by the relation

$$P_{emission} \propto \left(1 - \frac{\hbar\omega}{eV_b}\right) \times \frac{\rho_{total}}{\rho_0} \times \eta_{rad}, \quad (1)$$

where $V_b$ is the applied bias voltage, $\rho_0$ is the vacuum density of states, $\rho_{total}$ is the total LDOS, and $\eta_{rad}$ is the radiative efficiency of the antenna. Higher values of $\rho_{total}$ improve the overall tunneling efficiency, and a higher $\eta_{rad}$ increases the overall radiative efficiency of the emission spectrum.

To model the optical response of the device, we use the commercial Finite Difference Time Domain (FDTD) solver Lumerical FDTD™. The material data for silver is taken from the Johnson and Christy's database of optical constants [18], and the refractive index of SiO$_2$ is assumed to be non-dispersive with a value of $n_{barrier}$=1.456. A single electric dipole is placed at different positions within the gap, along the $x$ direction, to mimic the current flow within the device. Due to the finite lateral width of the MIM junction, the plasmon mode within the gap gets reflected at the end faces. This results in the formation of a standing wave between the two terminations, with the maximum of electric field at the edges [19]. In other words, a Fabry Pérot cavity is formed with highly confined gap plasmon modes within the barrier, which can be excited using a dipole source. The large LDOS within the gap leads to an enhanced emission as seen from eq. 1. The total LDOS at each position is calculated as $\rho_{total}=\rho_0 \times (P_{total}/P_0)$, where $P_{total}$ is the total power dissipated by a dipole in the presence of the structure, and $P_0$ is the power dissipated by a dipole of equal dipole moment in a homogeneous environment or vacuum [20]. The total LDOS as seen by the tunneling electrons is then determined as the average of respective LDOS at different positions. Similarly, the radiative LDOS is calculated as $\rho_{rad}=\rho_0 \times (P_{rad}/P_0)$, where $P_{rad}$ is the power radiated by the dipole in the far-field. The radiative efficiency, $\eta_{rad}$, is then calculated as the ratio of the radiated power, $P_{rad}$, to the total power, $P_{total}$. The nonlocal effects on gap plasmon resonance can be neglected, and all the calculations have been performed within the classical regime as the quantum effects are negligible for a tunnel barrier of width of 3 nm [21, 22]. By varying the $w$ and $h$ of the top

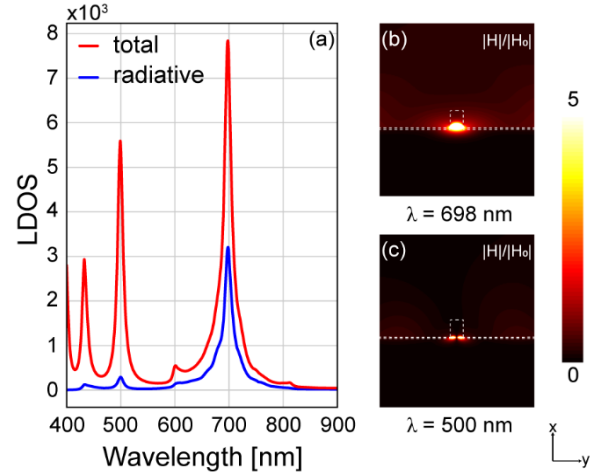

Figure 2. (a) Total (red) and radiative (blue) LDOS normalized to vacuum LDOS for $w$ = 35 nm, $h$ = 80 nm and $\Lambda$ = 400 nm. The LDOS shows an improvement of ~3 orders of magnitude over vacuum. (b) The H-field intensity profile of the mode at a wavelength of 698 nm. The mode is highly confined within the gap, and is magnetic dipole in nature with the dipole moment in $z$ direction. (c) Higher-order magnetic mode at 500 nm due to Fabry Pérot resonance.

 

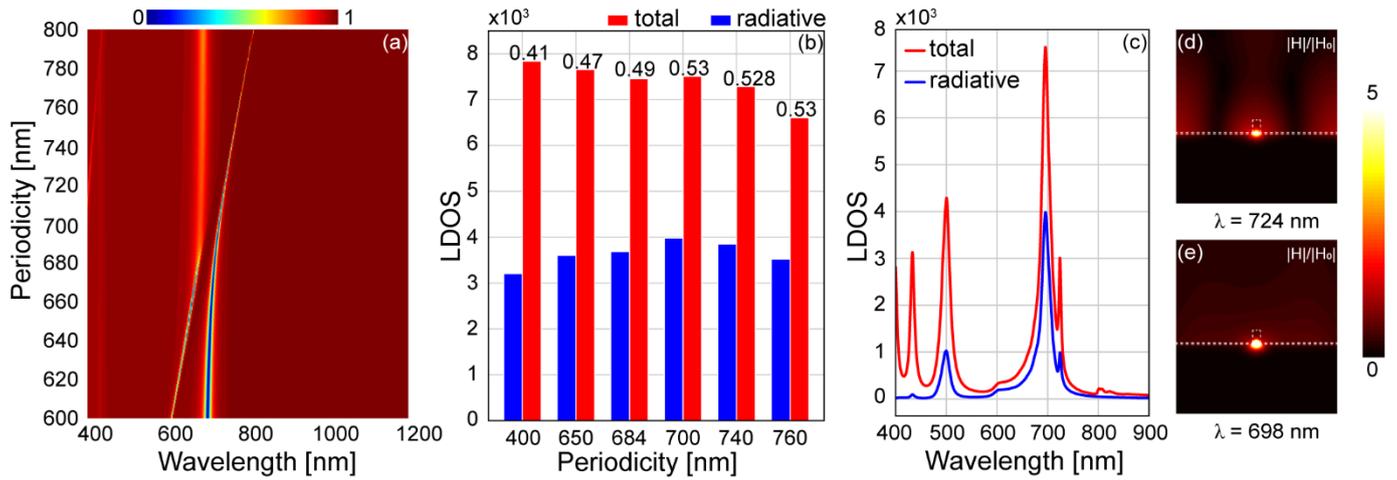

Figure 3. (a) Reflectivity as a function of periodicity and wavelength. The plot shows an anti-crossing between the SLR and gap plasmon mode at a periodicity of ~700 nm and a wavelength of 698 nm. (b) Comparison of radiative and total LDOS calculated for gap plasmon mode (λ = 698 nm) at different periodicities shows a higher radiative LDOS at a periodicity of 700 nm due to strong coupling between gap plasmon mode and lattice resonance. The radiative efficiency is depicted at the top of the bar for corresponding periodicities. (c) Total and radiative LDOS for the 700 nm periodicity structure. The two peaks at wavelength 698 nm and 724 nm agree well with the reflectivity calculations of (a), and indicate the coupling between gap plasmon mode and lattice resonance. (d, e) H-field profiles for the respective peaks. The peak at 724 nm shows the hybridization of gap plasmon and lattice resonance mode, with an enhanced emission in free-space.

electrode, and the periodicity, the overall radiative efficiency can be optimized to the desired wavelength range.

## 3. Results and Discussion

The normalized total and radiative LDOS for the structure with nano-strips of width 35 nm and height 80 nm is plotted in Fig. 2(a). The dimensions are optimized for the highest radiative efficiency for the fundamental gap plasmon mode for a periodicity of 400 nm. The optimal design is obtained by de-coupling the gap plasmon mode with the nano-strip antenna mode by varying the height of the top electrode. Though, intuitively, a coupling between the gap mode and an antenna mode increases the overall radiative efficiency, it also results in a decrease in the total LDOS leading to a lower inelastic electron tunneling rate [23]. Thus, the geometry is chosen such that we achieve a higher radiative efficiency without compromising the efficiency of IET. The total and radiative LDOS show a good ~3 orders of magnitude improvement over vacuum. The fundamental gap plasmon mode occurs at a wavelength of 698 nm, and the de-coupled antenna mode occurs at 600 nm. The H-field distribution in Fig. 2(b) shows that the fundamental mode is highly concentrated within the gap and is magnetic dipolar in nature with dipole moment in the z direction. The peaks at lower wavelengths correspond to higher-order magnetic dipolar modes of the Fabry Pérot cavity formed as seen in Fig. 2(c). Though the total LDOS for higher-order modes at 450 nm and 500 nm are comparable to the fundamental mode, they have a very low radiative LDOS leading to lower radiative efficiencies. Thus, the photon emission from these strucures can be attributed to the radiative out-coupling of the fundamental gap plasmon mode. The highest radiative LDOS of ~3200 is achieved for the fundamental mode at 698 nm, and the radiative efficiency for the structure is found to be ~0.41.

We now study the effects of varying periodicity on the radiative efficiency of the tunnel junction. The periodicity introduces an additional photonic mode, the surface lattice resonance (SLR), which results in coupling between the diffractive SLR mode and the gap plasmon mode, and can significantly increase the radiative out-coupling of the latter, thereby improving the overall radiative efficiency of the tunnel junction [24]. The SLR and the gap plasmon mode can be independently tuned by changing the periodicity, and thus bringing these modes into resonance. To explore the interplay between the gap plasmon mode and the SLR, reflectivity versus wavelength is calculated with varying periodicity for a plane wave incident at an angle normal to the surface. The corresponding data is plotted in Fig. 3(a). A dip at the longer wavelength corresponds to the gap plasmon mode, whereas the dip occurring at the shorter wavelength corresponds to the lattice resonance. At the gap plasmon resonance, the reflection dip goes to almost zero indicating that most of the incident energy of the plane wave is efficiently coupled to the gap plasmon mode. As the periodicity increases, the position of the gap plasmon mode remains relatively unchanged while the lattice resonance red-shifts. A further increase in periodicity results in an anti-crossing between the lattice resonance and the gap plasmon mode, which is consistent with experimental findings involving two-dimensional arrays of nanocubes [25]. The anti-crossing represents a region of strong coupling between the SLR and gap plasmon mode and occurs at a periodicity in the range of 690 nm – 700 nm, and a wavelength of 698 nm. There is a dramatic decrease in the intensity of the gap mode at the anti- crossing due to its interaction with the SLR. Beyond 700 nm, the intensity of gap plasmon resonance further reduces, indicating that it is weakly confined within the gap due to radiative out-coupling, and that the plane wave source is unable to excite the gap plasmon mode efficiently. The total LDOS for the gap plasmon in this region is lower and hence is not of interest to us. A second anti-crossing is also observed at lower wavelengths and higher periodicities due to a second-order lattice mode coupling to a higher-order Fabry-Pérot resonance.

We now determine the impact of varying periodicity on the respective local density of optical states and radiative efficiency of the structure. The total and radiative LDOS for the gap plasmon mode at different periodicities is plotted in Fig. 3(b). The data for 400 nm periodicity is included for comparison. The total LDOS remains more or less constant till the 700 nm periodicity, but decreases by ~1000 as the periodicity is increased to 760 nm. This is in agreement with the reflectivity calculations of Fig. 3(a) wherein the excitation of gap plasmon mode grows weaker beyond the coupling regime. Radiative LDOS, on the other hand, increases with increasing periodicity, peaks at 700 nm, and then gradually decreases. This is because a portion of the gap plasmon mode that is excited due to the inelastic tunneling of electrons decays as propagating SPs at the Ag-SiO$_2$ interface [17]. These SPs cannot out-couple to free-space radiation at lower periodicities due to the momentum mismatch between SPs and photons. Increasing the periodicity provides additional momentum in the form of lattice resonance and enables the out-coupling of SPs, thereby increasing the fraction of gap plasmon (total LDOS) converting to free-space photons (radiative LDOS). The $\eta_{rad}$ for each periodicity is calculated and mentioned at the top of the corresponding bars within the graph. Highest radiative efficiency of ~0.53 occurs at the periodicity of 700 nm. Beyond 700 nm, though the overall $\eta_{rad}$ remains more or less constant, there is an overall decrease in the total and radiative LDOS. Thus, the 700 nm periodicity can be considered as an optimum value for observing a strong interaction between the SLR and the gap plasmon mode without compromising on the efficiency of IET. On comparing the radiative efficiencies for the 400 nm and 700 nm periodicity, there is an increase of ~ 30% for the 700 nm periodicity, diffractively coupled, tunnel junction. The total and radiative LDOS at the optimum periodicity of 700 nm are shown in Fig. 3(c). The gap plasmon resonance at 698 nm wavelength is followed by another peak at a wavelength of 724 nm. The peaks are in excellent agreement with the corresponding reflectivity dips for a periodicity of 700 nm shown in Fig. 3(a), indicating the hybridization of the gap plasmon and lattice resonance modes. The corresponding H-field profiles for the two peaks are shown in Fig. 3(d, e). There is significant emission into the top air half-space as compared to the 400 nm periodicity (Fig. 3(d)), indicating the efficient out-coupling of gap plasmon to free-space radiation. The corresponding spectrum for different applied bias voltages for periodicity of 700 nm is calculated using eq. (1) and plotted in Fig. 4. The spectral profile follows the shape of the total LDOS, as seen from Fig. 3(c). The spectral shape agrees well with the previously reported data [1, 5], and follows the quantum cutoff condition wherein the mode is excited only when the applied bias exceeds the gap plasmon frequency, $|eV_b| = \hbar\omega$ [1]. For a bias voltage of 1.8 V, gap plasmon mode is weakly excited, and the spectrum shows no peaks. As the bias voltage is increased, there is a gradual rise in the overall emission intensity with a relative increase in intensity at the shorter wavelength range.

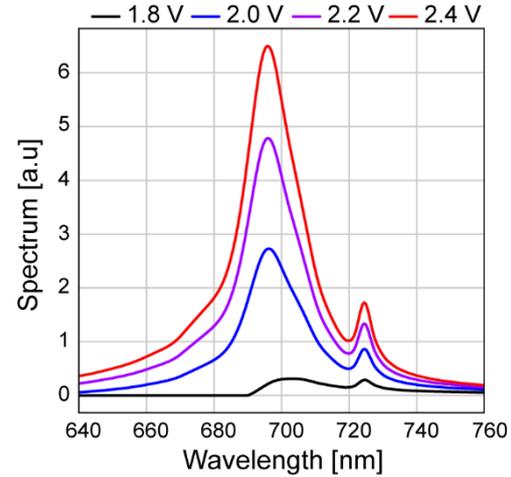

Figure 4. Spectrum of emitted radiation at different applied bias voltages. As the applied bias is increased, a well-defined peak is formed due to the gap plasmon excitation. The spectra also defines a high energy cut-off depending on the applied bias, $|eV_b| = \hbar\omega$.

In conclusion, we numerically investigated the light generation through inelastic electron tunneling in a periodic Ag-SiO$_2$-Ag tunnel junction with a 3 nm barrier width. We showed that the inelastic tunneling current efficiently excites the gap plasmon mode with high local density of optical states within the barrier. The dimensions of the top electrode were optimized for efficient out-coupling of the fundamental gap plasmon mode to free-space photons at the desired wavelength. To further enhance the radiative efficiency, we explored the interplay between gap plasmon and lattice resonance by varying the periodicity of the structure. An anti-crossing between the two modes was observed at a periodicity of 700 nm. The hybridization between the lattice resonance and gap plasmon mode at the anti-crossing led to an enhancement in the radiative efficiency by ~30% as compared to the structure without the lattice resonance. This improvement is due to the direct out-coupling of propagating SPs to free-space photons mediated by the grating. The emitted spectrum was found to be dependent on the total LDOS of the gap plasmon mode, and hence can be tuned to the desired wavelength range, from visible to near IR, by designing the geometrical and material parameters of the device accordingly. The emission efficiency may be further improved by, for example, reducing the gap size or using a high-index dielectric as an insulator so that the devices can be readily integrated for on-chip photonic and plasmonic applications.

## Acknowledgements

This work was supported by the Science and Engineering Research Board (SERB) through Ramanujan fellowship (SB/S2/RJN-007/2017) and Early Career Research (ECR/2018/002452) grants. SK and JT thank the Ministry of Human Resource Development (MHRD), Govt. of India for the research fellowship to undertake Ph.D. study.